# Electric field enhancement of pool boiling of dielectric fluids on pillar-structured surfaces: A lattice Boltzmann study

Wanxin Li, Qing Li*, Yue Yu, and Shi Tang

*School of Energy Science and Engineering, Central South University, Changsha 410083, China*

*Corresponding author: qingli@csu.edu.cn

## Abstract

In this paper, by using a phase-change lattice Boltzmann (LB) model coupled with an electric field model, we numerically investigate the performance and enhancement mechanism of pool boiling of dielectric fluids on pillar-structured surfaces under an electric field. The numerical investigation reveals that applying an electric field causes both positive and negative influences on the pool boiling of dielectric fluids on pillar-structured surfaces. It is found that, under the action of an electric field, the electric force prevents the bubbles nucleated in the channels from crossing the edges of the pillar tops. On the one hand, such an effect results in the bubble coalescence in the channels and blocks the paths of liquid supply for the channels, which leads to the deterioration of pool boiling in the medium-superheat regime. On the other hand, it prevents the coalescence between the bubbles in the channels and those on the pillar tops, which suppresses the formation of a continuous vapor film and therefore delays the occurrence of boiling crisis. Meanwhile, the electric force can promote the departure of the bubbles on the pillar tops. Accordingly, the critical heat flux (CHF) can be improved. Based on the revealed mechanism, wettability-modified regions are applied to the pillar tops for further enhancing the boiling heat transfer. It is shown that the boiling performance on pillar-structured surfaces can be enhanced synergistically with the CHF being increased by imposing an electric field and the maximum heat transfer coefficient being improved by applying mixed wettability to the pillar-structured surfaces.

# 1. Introduction

Boiling heat transfer, which is one of the most efficient modes for heat dissipation, has been widely employed in daily life and in many industrial fields, such as nuclear power reactors [1], heat exchangers [2], refrigeration [3], and electronic devices [4]. Compared with single-phase thermal convection, boiling heat transfer can provide a much higher heat transfer coefficient (HTC) because of owning a large latent heat consumption of liquid-vapor phase change. However, the capacity of boiling heat transfer is limited by the so-called critical heat flux (CHF), beyond which the heating surface may be covered by a continuous vapor film and the boiling crisis occurs [5]. Therefore, an enormous number of experimental and numerical studies have been conducted to explore the methods for enhancing boiling heat transfer with the CHF and/or the HTC being increased.

Generally, the available methods for the enhancement of boiling heat transfer can be divided into two main categories, i.e., passive methods and active ones [6]. The passive methods are mainly relied on the modification of fluid properties [7], surface wettability [8], micro/nano-structures [9], etc. With the rapid development of surface modification and micromachining technologies, the boiling enhancement by using the modification of surface structure and/or surface wettability has been the focus of a significant number of studies [10-14]. In the literature, it has been revealed that a hydrophobic surface with poor wettability usually enhances the boiling heat transfer with a higher HTC [15], while a hydrophilic surface owning higher liquid affinity ability can delay the occurrence of the CHF [16]. Therefore, in recent years some heating surfaces with mixed wettability have been fabricated to enhance boiling heat transfer [14, 17-19]. Moreover, surface structure also plays a very important role in boiling heat transfer. In the past decades, various types of structured surfaces have been proposed and fabricated to enhance boiling performance, such as surfaces with microcavities [20], micropillars [8, 21], and micro-nano composited structures [22-24]. Compared with a plain surface, a structured surface can offer more activated nucleation sites for enhancing boiling heat transfer [9], and the CHF can also be improved by the capillary wicking supplied by a roughly surface structure [25]. Different from the passive methods, the active techniques of enhancing boiling heat transfer usually employ external power such as the magnetic and electric fields. For example,

Ogata and Yabe [26] experimentally found that applying an electric field can increase the number of bubbles and decrease the diameter of the departed bubbles. Dong *et al*. [27] showed that the bubbles under an electric field are elongated vertically by the electric force and large bubbles may be broken into small ones while departing from the heating surface. Hence, with such distinctive features of bubble dynamics under an electric field, the CHF can be increased relative to that without imposing the electric field [28].

In order to take the advantages of both passive and active techniques, the influence of surface structure and the electric field effect have been combined for enhancing boiling heat transfer in recent years. By imposing an external electric field, Kano [29] has carried out boiling experiments on structured surfaces with a micro-sized electrode at $-5$ kV/mm being applied. They found that the combined effects of electric field and surface structure can increase the CHF with the value being much larger than that of a smooth surface without the electric field. Similarly, Garivalis *et al*. [30] have also studied such combined effects under microgravity. They found that the electric field can create gravity-mimicking body forces. However, for the bubbles nucleated from the bottom of the surface structures, the growth and detachment of these bubbles may be inhibited by the electric force, and therefore the boiling heat transfer may be deteriorated. For instance, Quan *et al*. [31] experimentally explored the boiling performance on rib surfaces under mesh-structured electrodes. They found that the boiling heat transfer is deteriorated in the medium-superheat region by an electric field. Actually, such a phenomenon of heat transfer deterioration by an electric field also exists in the experimental study of Kano *et al.* [29]. Moreover, Liu *et al*. [32] also experimentally found that applying an electric field is not always beneficial to the CHF enhancement of micro-pin-fins.

With the rapid development of high-performance computational technology, computational fluid dynamics (CFD) has become more and more important in studying multiphase flow and phase-change heat transfer [33]. Compared with experiments, numerical simulations can provide more details of boiling processes for promoting the understanding of the mechanism of boiling heat transfer [34], such as the temperature contours, the velocity vectors, and the distributions of the electric force. In recent years, the lattice Boltzmann (LB) method, which originated from the lattice-gas automata method, has been extensively employed to simulate boiling phenomena [35-42]. Specifically, this method has also been applied to investigate the influences of an electric field on boiling heat transfer. Feng *et al*. [43, 44] employed an electric field model coupled with the phase-change LB model proposed by Li *et al*. [36] to study the boiling heat transfer on flat surfaces under an electric field. They found that the electric field can enhance the boiling performance in the fully developed nucleate boiling regime, and increasing the electric field intensity can decrease the bubble departure diameter and the bubble release period. Recently, Yao *et al*. [45] also investigated the effects of an electric field on bubble dynamics and found that the electric force acting on the liquid-vapor interface plays an important role in the detachment of bubbles. However, these LB simulations of pool boiling under an electric field were focused on flat surfaces, and up to now no studies have been reported on investigating the performance of pool boiling on structured surfaces under an electric field by using the LB method.

Moreover, it should be noted that applying an electric field may deteriorate the boiling heat transfer according to the aforementioned experimental studies [29, 31]. Therefore, the pool boiling of dielectric fluids on structured surfaces under an electric field needs to be further enhanced. Based on this consideration, in the present study we firstly employ the LB method to investigate the pool boiling performance of dielectric fluids on pillar-structured surfaces in the presence of an electric field, with a focus being placed on numerically revealing the associated boiling mechanism. Subsequently, with the guided orientation of the revealed mechanism, the pool boiling on pillar-structured surfaces is enhanced by combining the electric field effect and the effect of mixed wettability. It is shown that the pool boiling on pillar-structured surfaces can be enhanced synergistically with the CHF being increased by imposing an electric field and the maximum HTC being improved by applying mixed wettability to the pillar-structured surfaces. The rest of the present paper is organized as follows. A phase-change LB model in the presence of an electric field is briefly introduced in Sec. 2. Numerical investigation of the performance and the associated enhancement mechanism of pool boiling on pillar-structured surfaces under an electric field is conducted in Sec. 3. Moreover, the boiling enhancement by combining the electric field effect and the effect of mixed wettability is also studied in Sec. 3. Finally, a brief summary is provided in Sec. 4.

## 2. Numerical method

Similar to the studies in Refs. [43, 44], the present work also employs an electric field model coupled with the phase-change LB model proposed by Li *et al.* [33, 36] to simulate the boiling processes under an electric field. By using a multi-relaxation-time (MRT) collision operator, the LB equation that governs the evolution of the density distribution function can be written as follows [46-48]:

$$f_\alpha(\mathbf{x}+\mathbf{e}_\alpha\delta_t,\ t+\delta_t) = f_\alpha(\mathbf{x},\ t) - \overline{\Lambda}_{\alpha\beta}\left(f_\beta - f_\beta^{eq}\right)\Big|_{(\mathbf{x},\ t)} + \delta_t\left(G_\alpha - 0.5\overline{\Lambda}_{\alpha\beta}G_\beta\right)\Big|_{(\mathbf{x},\ t)}, \tag{1}$$

where $f_\alpha$ and $f_\alpha^{eq}$ are the density distribution function and the equilibrium density distribution function, respectively, $\mathbf{x}$ is the spatial position, $\mathbf{e}_\alpha$ is the discrete velocity, $t$ is the time, $\delta_t$ is the time step, $G_\alpha$ is the forcing term in the discrete space, and $\overline{\Lambda}_{\alpha\beta} = \left(\mathbf{M}^{-1}\mathbf{\Lambda}\mathbf{M}\right)\Big|_{\alpha\beta}$ is the MRT collision operator, in which $\mathbf{M}$ is a transformation matrix and $\mathbf{\Lambda} = \text{diag}\left(\tau_\rho^{-1},\ \tau_e^{-1},\ \tau_\varsigma^{-1},\ \tau_j^{-1},\ \tau_q^{-1},\ \tau_j^{-1},\ \tau_q^{-1},\ \tau_\upsilon^{-1},\ \tau_\upsilon^{-1}\right)$ is a diagonal matrix containing the relaxation times for the D2Q9 lattice model.

Using the transformation matrix $\mathbf{M}$, the right-hand side of Eq. (1) can be mapped onto the moment space, i.e.,

$$\mathbf{m}^* = \mathbf{m} - \mathbf{\Lambda}\left(\mathbf{m} - \mathbf{m}^{eq}\right) + \delta_t\left(\mathbf{I} - \frac{\mathbf{\Lambda}}{2}\right)\mathbf{S}, \tag{2}$$

where $\mathbf{m} = \mathbf{M}\mathbf{f}$, $\mathbf{m}^{eq} = \mathbf{M}\mathbf{f}^{eq}$, $\mathbf{I}$ is the unit matrix, and $\mathbf{S} = \mathbf{M}\mathbf{G}$ is the forcing term in the moment space. The details of the equilibria $\mathbf{m}^{eq}$ and the forcing term $\mathbf{S}$ can be found in Ref. [36]. In the LB community, Eq. (2) is usually called the collision step. By applying an inverse matrix $\mathbf{M}^{-1}$ of the transformation matrix, $\mathbf{m}^*$ obtained by Eq. (2) can be transformed back to the discrete velocity space with $\mathbf{f}^* = \mathbf{M}^{-1}\mathbf{m}^*$, and the streaming step is given by

$$f_\alpha(\mathbf{x}+\mathbf{e}_\alpha\delta_t,\ t+\delta_t) = f_\alpha^*(\mathbf{x},\ t). \tag{3}$$

Meanwhile, the macroscopic density and velocity can be calculated by

$$\rho = \sum_\alpha f_\alpha, \quad \rho \mathbf{u} = \sum_\alpha \mathbf{e}_\alpha f_\alpha + \frac{\delta_t}{2}\mathbf{F}, \tag{4}$$

where $\mathbf{F}$ is the total force acting on the system, which includes the intermolecular interaction force $\mathbf{F}_m$, the buoyancy force $\mathbf{F}_b$, and the electric force $\mathbf{F}_e$.

The intermolecular interaction force for single-component multiphase systems is defined as [49]:

$$\mathbf{F}_m = -G\psi(\mathbf{x})\sum_\alpha w_\alpha \psi(\mathbf{x}+\mathbf{e}_\alpha \delta_t)\mathbf{e}_\alpha, \tag{5}$$

where $G$ is a parameter that tunes the strength of the interaction force, $\psi(\mathbf{x})$ is the pseudopotential at the site $\mathbf{x}$, and $w_\alpha$ are the weights given by $w_{1-4}=1/3$ and $w_{5-8}=1/12$, respectively. The buoyancy force is given by $\mathbf{F}_b = (\rho - \rho_{ave})\mathbf{g}$, where $\rho_{ave}$ is the average density in the computational domain and $\mathbf{g}$ is the gravitational acceleration. The pseudopotential in Eq. (5) is taken as

$$\psi(\mathbf{x}) = \sqrt{2(p_{EOS} - \rho c_s^2)/Gc^2}, \tag{6}$$

where $p_{EOS}$ is the non-ideal equation of state and $c_s = c/\sqrt{3}$ is the lattice sound speed, in which $c=1$ is the lattice speed. In this work, the Peng-Robinson equation of state is adopted [50]

$$p_{EOS} = \frac{\rho RT}{1-b\rho} - \frac{a\varphi(T)\rho^2}{1+2b\rho - b^2\rho^2}, \tag{7}$$

where $\varphi(T)$ is given by

$$\varphi(T) = \left[1 + (0.37464 + 1.54226\omega - 0.26992\omega^2)(1 - \sqrt{T/T_c})\right]^2, \tag{8}$$

in which $\omega = 3.44$ is the acentric factor, $a = 0.45724 R^2 T_c^2/p_c$, and $b = 0.0778 R T_c/p_c$, where $T_c$ and $p_c$ are the critical temperature and the critical pressure, respectively. Following Ref. [36], the parameters in the present LB simulations are chosen as $a = 3/49$, $b = 2/21$, and $R = 1$.

By neglecting viscous heat dissipation, the temperature field of non-ideal fluids is governed by [36]

$$\partial_t T = \mathbf{u}\cdot\nabla T = \frac{1}{\rho c_V}\nabla\cdot(\lambda \nabla T) - \frac{T}{\rho c_V}\left(\frac{\partial p_{EOS}}{\partial T}\right)_\rho \nabla\cdot\mathbf{u}, \tag{9}$$

where $c_V$ is the specific heat at constant volume and $\lambda$ is the thermal conductivity taken as

$$\lambda = \lambda_v \frac{\rho_l - \rho}{\rho_l - \rho_v} + \lambda_l \frac{\rho - \rho_v}{\rho_l - \rho_v}, \tag{10}$$

where $\lambda_v = \rho_v c_V \chi_v$ is the thermal conductivity of the vapor phase, in which $\chi_v$ is the thermal diffusivity. In the present work, the temperature equation given by Eq. (9) is solved by using the classical fourth-order Runge-Kutta scheme for the time discretization and the isotropic central finite-difference schemes for the spatial discretization. The details of these schemes can be found in Refs. [36, 38].

Based on the electro-hydrodynamics theory [51], the electric force $\mathbf{F}_e$ can be written as follows [52]:

$$\mathbf{F}_e = q_e \mathbf{E} - \frac{1}{2} \mathbf{E} \cdot \mathbf{E} \nabla \varepsilon + \frac{1}{2} \nabla \left[ \rho \frac{\partial \varepsilon}{\partial \rho} \mathbf{E} \cdot \mathbf{E} \right], \tag{11}$$

where $q_e$ is the free charge density, $\varepsilon$ is the dielectric permittivity of the fluid, and $\mathbf{E}$ is the electric field intensity. On the right-hand side of Eq. (11), the first term is the Coulomb force, the second term is the dielectric electrophoretic force, and the third term is the electrostriction force. For dielectric and incompressible fluids, the Coulomb force and the electrostriction force can be neglected [52]. Therefore, Eq. (11) can be rewritten as

$$\mathbf{F}_e = -\frac{1}{2} \mathbf{E} \cdot \mathbf{E} \nabla \varepsilon. \tag{12}$$

As the perfect dielectric medium has no free charge, the intensity of the electric field can be solved by $\nabla \cdot (\varepsilon \mathbf{E}) = 0$. In the absence of the magneto-induced effect, the curl of the electric field is zero and the electric field intensity is the gradient of the electric potential $U$, i.e., $\mathbf{E} = -\nabla U$ [52]. Hence, the equation that governs the electric field can be expressed as follows:

$$\nabla \cdot (\varepsilon \nabla U) = 0. \tag{13}$$

In order to solve Eq. (13), the following LB equation is applied according to Refs. [43, 53]:

$$g_\alpha (\mathbf{x} + \mathbf{e}_\alpha \delta_t, t + \delta_t) - g_\alpha (\mathbf{x}, t) = -\frac{1}{\tau_s} \left[ g_\alpha (\mathbf{x}, t) - g_\alpha^{eq} (\mathbf{x}, t) \right], \tag{14}$$

where $\tau_s = 3\varepsilon + 0.5$ is the relaxation time, $g_\alpha$ and $g_\alpha^{eq}$ are the distribution function and the equilibrium distribution function of the electric potential, respectively. The equilibrium distribution function of the electric potential is given by $g_\alpha^{eq} = \omega_\alpha U$, in which the weights $\omega_\alpha$ are chosen as $\omega_{1-4} = 1/9$ and $\omega_{5-8} = 1/36$, respectively. Then, the electric potential is obtained as follows:

$$U = \sum_{\alpha} g_{\alpha}(\mathbf{x}, t). \tag{15}$$

In the present work the dielectric permittivity of the fluid is expressed by [45]

$$\varepsilon = \varepsilon_v \frac{\rho_l - \rho}{\rho_l - \rho_v} + \varepsilon_l \frac{\rho - \rho_v}{\rho_l - \rho_v}, \tag{16}$$

where $\varepsilon_v$ and $\varepsilon_l$ are the dielectric permittivity of the vapor and liquid phases, respectively. After obtaining the chemical potential from Eq. (15), the electric field intensity can be calculated according to $\mathbf{E} = -\nabla U$. Then the electric force can be obtained via Eq. (12).

## 3. Numerical results and discussion

### 3.1. Boiling performance on pillar-structured surfaces under an electric field

#### *3.1.1. Simulation setup*

Figure 1 illustrates the physical model of boiling simulations on pillar-structured surfaces under an electric field. In our simulations, the computational domain is taken as $L_x \times L_y = 450 \text{ l.u.} \times 300 \text{ l.u.}$, where l.u. represents the lattice units in the LB method. The pillar-structured surface is located at the bottom of the computational domain. The height of each pillar is fixed at $H = 40 \text{ l.u.}$ and the center-to-center spacing between neighboring pillars is taken as $s = 150 \text{ l.u.}$ The heating surface is a homogeneous hydrophilic surface and its intrinsic contact angle is $\theta_{\text{phi}} \approx 36.6°$, whose numerical implementation scheme can be found in Ref. [54]. Initially, the computational domain is occupied by a saturated dielectric liquid ($0 \text{ l.u.} \leq y \leq 200 \text{ l.u.}$) below its saturated vapor ($200 \text{ l.u.} < y \leq L_y$). The initial temperature of the computational domain is set to be $T_s = 0.86 T_c$, which corresponds to the coexistence densities $\rho_l \approx 6.5$ and $\rho_v \approx 0.38$, respectively. The temperature of the heating surface is taken as $T_b = T_s + \Delta T$, in which $\Delta T$ is the wall superheat. Following Ref. [36], the specific heat at constant volume in the present LB simulations is chosen as $c_V = 6.0$. The thermal diffusivity of the vapor phase is taken as $\chi_v = 0.01$ and the ratio of the thermal conductivity between the liquid and vapor is chosen as $\lambda_l / \lambda_v = 15$. Besides, the gravitational acceleration is given by $\mathbf{g} = (0, -3.0 \times 10^{-5})$.

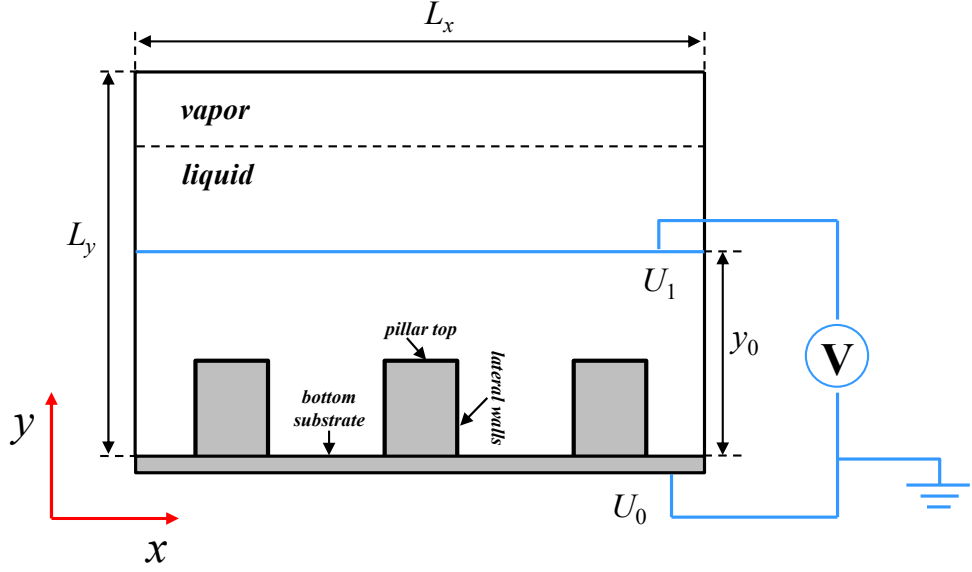

**Fig. 1.** Physical model of boiling simulations on pillar-structured surfaces under an electric field.

As for the electric field settings shown in Fig. 1, the heating surface is taken as the low electrode that links with the ground ($U_0 = 0$) and the high electrode with the electric potential of $U_1$ is located above the heating surface. The distance between the low electrode and the bottom substrate is chosen as $y_0 = 80$ l.u. Thus, the initially imposed electric field intensity is given by $E_0 = (U_1 - U_0)/y_0$. In the present work, the periodic boundary condition is applied in the horizontal direction for both the flow and electric fields. The Zou-He boundary scheme [55] and its modified mass-conservative boundary scheme [56] are used at the heating surface so as to satisfy the no-slip boundary condition of the flow field, while for the electric field the non-equilibrium extrapolation scheme [43, 57] is employed at the low and high electrodes. In addition, the dielectric permittivity of the vapor phase is taken as $\varepsilon_v = 0.1$ and the ratio is chosen as $\varepsilon_l/\varepsilon_v = 2.236$ following Refs. [43, 44]. Note that, the quantities in the present study are all taken in lattice units and the conversion between the lattice units and the physical units can be found in Ref. [58]. Besides, the simulations are performed without the temperature solver and the electric field solver during the first 1000 time steps.

### 3.1.2. Boiling curves and HTCs

Firstly, the boiling performance on the pillar-structured surface under an electric field is investigated, including the boiling curves and HTCs. Figure 2(a) illustrates the boiling curves of pool boing on the pillar-structured surface with and without an electric field when the pillar width is taken as $W = 60$ l.u. The corresponding HTCs are presented in Fig. 2(b). In the present work, the heat flux is normalized as $q^* = q_{ave} \Delta x/(\lambda_v T_c)$, where $q_{ave}$ is the time and spatial average of the local transient heat flux during the boiling process of $t \in \left[10^3 \delta_t, 5 \times 10^4 \delta_t\right]$, $\Delta x$ is the grid spacing, $\lambda_v$ is the thermal conductivity of the vapor phase, and $T_c$ is the critical temperature. The normalized HTC is given by $h^* = q^* T_c/\Delta T$. As seen in Fig. 2(a), the boiling curves show the variations of the normalized heat flux with increasing the wall superheat from the onset of the nucleate boiling (ONB) at $\Delta T \approx 0.0121$.

From Fig. 2(a) we can see that the boiling curves can be generally classified into three regimes, i.e., regimes I, II, and III, respectively. In the regime I, the heat fluxes of the cases under an electric field ($E_0 = 0.20$ and 0.25) are higher than that of the case without the electric field ($E_0 = 0$) and the heat flux increases with the increase of the electric field intensity under the same wall superheat. However, when the boiling enters the medium-superheat regime with $0.0148 < \Delta T < 0.0186$ (i.e., the regime II), the boiling performance is seemingly deteriorated by the electric field. As seen in Fig. 2(a), in the regime II the heat fluxes of the cases under the electric field are lower than that of the case without imposing the electric field. As for the regime III, when the wall superheat continues to increase, the heat flux of the case without the electric field decreases dramatically because of the appearance of the transition boiling. However, for the cases under the electric field the heat fluxes keep increasing until the CHF occurs, which means that the occurrence of boiling crisis is delayed owing to the influence of the electric field. Correspondingly, Fig. 2(b) also shows that the pool boiling HTC on the pillar-structured surface is deteriorated by the electric field in the regime II, although it is enhanced in the regimes I and III. At the wall superheat of $\Delta T \approx 0.0169$, which corresponds to the CHF and HTC$_{max}$ of the case without the electric field, the HTC is reduced by 13.3% when the electric field intensity increases from $E_0 = 0$ to $0.25$. To sum up, the boiling crisis can be delayed by applying an electric field and the CHF can be enhanced when an appropriate value of the electric field intensity is imposed, but the HTC is obviously deteriorated in the medium-superheat regime (i.e., the regime II).

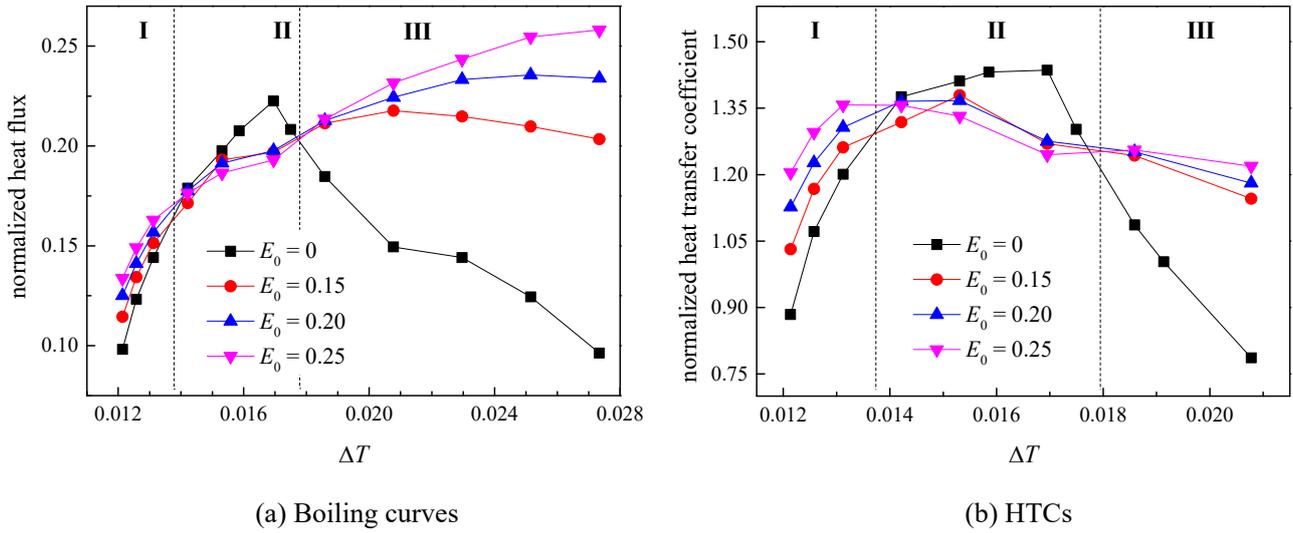

(a) Boiling curves  (b) HTCs

**Fig. 2.** Boiling curves and HTCs on the pillar-structured surface under an electric field ($W = 60$ l.u.).

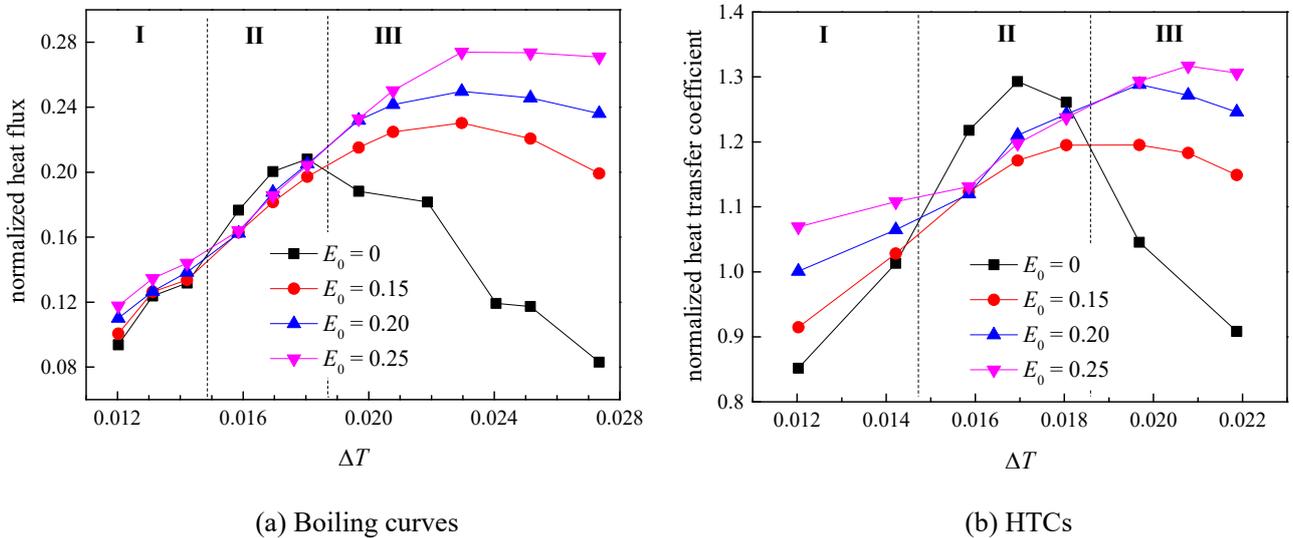

(a) Boiling curves  (b) HTCs

**Fig. 3.** Boiling curves and HTCs on the pillar-structured surface under an electric field ($W = 100$ l.u.).

Figure 3 depicts the boiling curves and HTCs of pool boiling on the pillar-structured surface with the pillar width of $W = 100$ l.u. Similar trends can be observed in Figs. 3(a) and 3(b) in the boiling curves and HTCs. As seen in the figures, the boiling curves and HTCs can also be divided into three regimes by judging whether the electric field effect can enhance the boiling performance or not. Specifically, the CHF can be increased by 31.7% when the electric field with $E_0 = 0.25$ is imposed. However, Fig. 3(b) also shows that the HTC in the regime II is deteriorated by the electric field, which further confirms that the CHF of pool boiling on the pillar-structured surface can be enhanced by applying an electric field, but the boiling heat transfer may be suppressed in the medium-superheat regime. To figure out the rationale behind these phenomena, the associated boiling mechanism will be investigated and discussed in the following section.

**3.2. Boiling mechanism on pillar-structured surfaces under an electric field**

In this section, the boiling mechanism of pool boiling on the pillar-structured surface under an electric field is analyzed, with a focus being placed on why the boiling heat transfer is enhanced in the regimes I and III, but deteriorated in the regime II. For convenience, the boiling simulations with $W = 60$ l.u. under the electric field are employed for analysis.

Firstly, the boiling heat transfer enhancement mechanism in the regime I is investigated. Figure 4 displays the temperature fields with the velocity vectors during the boiling processes at $\Delta T \approx 0.0126$ in the cases of $E_0 = 0$ and $E_0 = 0.25$, respectively. Note that, the displayed domain with one pillar is chosen here since in our simulations the periodic boundary conditions is used in the $x$ direction. The local normalized heat flux distributions of the two cases along the left-side wall, the pillar top, and the right-side wall of the pillar are compared in Fig. 4(c). From Figs. 4(a-i) and 4(b-i) we can see that at the instant of $t = 2000\delta_t$ the heat transfer is mainly dominant by the thermal conduction and the heat from the bottom surface has been barely diffused to the fluid. Therefore, it can be seen that the temperature distributions in the cases of $E_0 = 0$ and $E_0 = 0.25$ are almost identical and the local normalized heat fluxes of these two cases are basically consistent, as shown in Fig. 4(c-i).

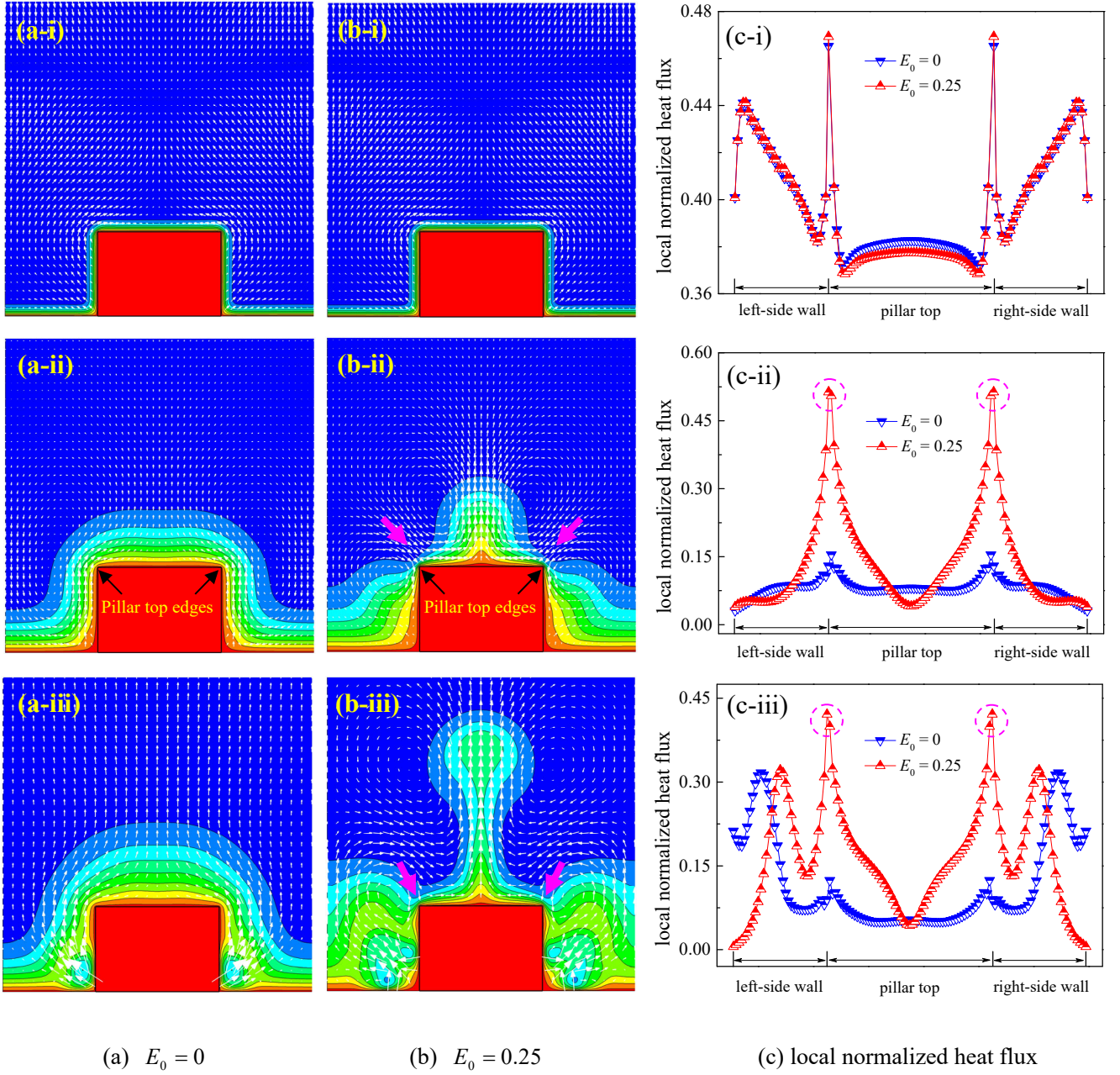

(a) $E_0 = 0$  (b) $E_0 = 0.25$  (c) local normalized heat flux

**Fig. 4.** Comparisons of the temperature fields with the velocity vectors at $\Delta T \approx 0.0126$ between the cases of (a) $E_0 = 0$ and (b) $E_0 = 0.25$. (c) The corresponding comparison of the local normalized heat flux distribution along the left-side wall, the pillar top, and the right-side wall of the pillar. From top to bottom: $t = 2000\delta_t$, $20000\delta_t$, and $30000\delta_t$, respectively.

However, at $t = 20000\delta_t$, the temperature fields of the two cases exhibit significant differences, as seen in Figs. 4(a-ii) and 4(b-ii). Specifically, the heat transfer in Fig. 4(a-ii) is still dominant by the thermal conduction, while the electroconvection caused by the electric field effect can be found in Fig. 4(b-ii), which can enhance the heat transfer. Particularly, from Fig. 4(b-ii) it can be seen that the thermal boundary layer is very thin around the edges of the pillar top when the electric field with $E_0 = 0.25$ is imposed, which corresponds to very large temperature gradients. Thus, the heat transfer around the edges of the pillar top is much stronger than that in other regions, as seen from Fig. 4(c-ii), which displays the distribution of the local normalized heat flux along the heating surface at $t = 20000\delta_t$. Similar phenomena can also be observed in Figs. 4(b-iii) and 4(c-iii) for the instant of $t = 30000\delta_t$, at which small bubbles have nucleated in the channels. Owing to the electroconvection caused by the electric field effect, the heat transfer in Fig. 4(b-iii) is also much stronger than that in Fig. 4(a-iii), as illustrated by the local normalized heat flux distributions in Fig. 4(c-iii).

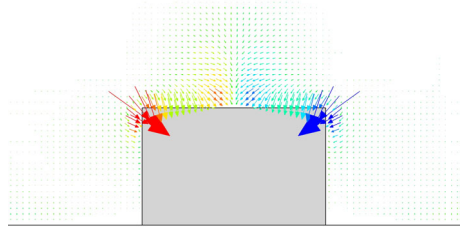

**Fig. 5.** Local distribution of the electric force in the case of $\Delta T \approx 0.0126$ and $E_0 = 0.25$ at $t = 20000\delta_t$.

To further illustrate the electric field effect in the regime I, the local distribution of the electric force around the pillar in the case of $\Delta T \approx 0.0126$ and $E_0 = 0.25$ at $t = 20000\delta_t$ is displayed in Fig. 5, which corresponds to the instant shown in Fig. 4(b-ii). As seen in Fig. 5, the electric force is very large around the edges of the pillar top. More importantly, the electric force points towards the heating surface, which means that the bulk liquid near the edges of the pillar top with low temperature will be driven to cool the heating surface under the action of the electric force. The velocity vectors denoted by the white arrows in Figs. 4(b-ii) and 4(b-iii) confirm such an analysis. This is also the reason why the heat transfer around the edges of the pillar top is much stronger than that in other regions.

Now attention turns to the boiling mechanism in the regime II. Taking the boiling simulations at $\Delta T \approx 0.0169$ as an example, which corresponds to the CHF on the pillar-structured surface without the electric field (see Fig. 2(a)), Fig. 6 displays some snapshots of the boiling processes at $\Delta T \approx 0.0169$ in the cases of $E_0 = 0$ and $E_0 = 0.15$, respectively. From Fig. 6(a) we can see that, for the case without the electric field ($E_0 = 0$), the bubbles nucleated in the channels gradually grow along the lateral walls of the pillars and then coalesce with the bubbles nucleated on the pillar tops. Subsequently, the coalesced bubbles are detached from the heating surface. However, when imposing the electric field ($E_0 = 0.15$), the bubbles nucleated in the channels will not coalesce with the bubbles on the pillar tops. In contrast, they will merge with their adjacent bubbles in each channel. As a result, the channels may be covered by the coalesced bubbles and the paths for liquid supply may be blocked, as seen in the right-hand panel of Fig. 6(b).

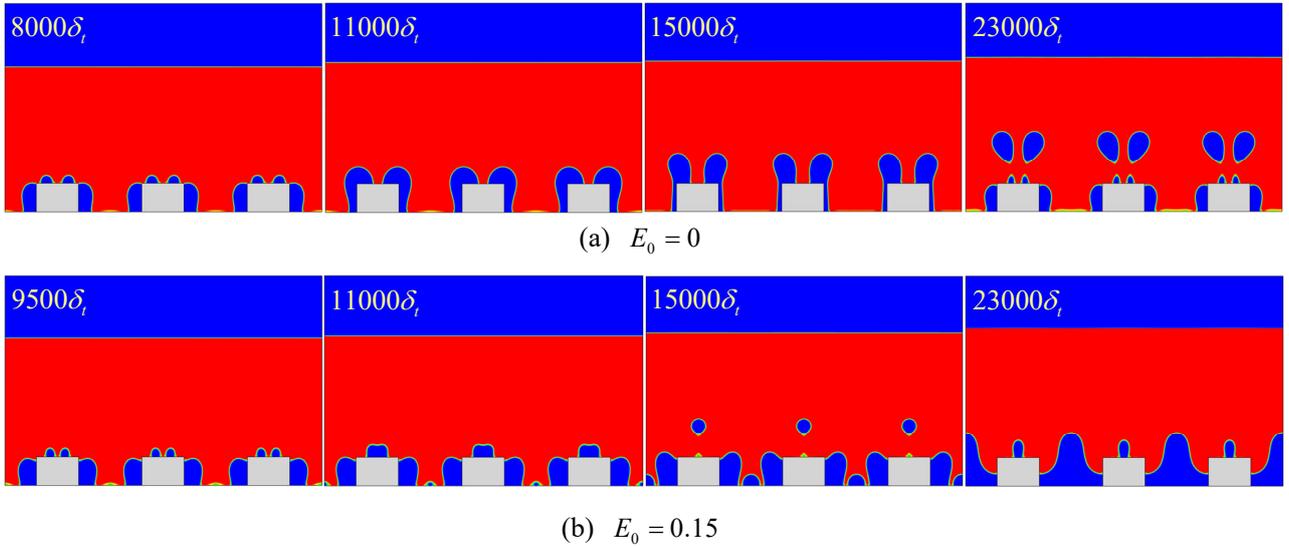

**Fig. 6.** Snapshots of the boiling processes on the pillar-structured surface at $\Delta T \approx 0.0169$ in the cases of (a) $E_0 = 0$ and (b) $E_0 = 0.15$, respectively.

To reveal the rationale behind the aforementioned phenomenon, Fig. 7 presents the local distributions of the electric force in the case of $\Delta T \approx 0.0169$ and $E_0 = 0.15$ at $t = 11000\delta_t$ and $t = 15000\delta_t$, respectively. As seen in Fig. 7, the electric force mainly acts on the liquid-vapor interface and basically points to the interior of the bubbles along the normal direction of the interface. Moreover, it can be clearly seen that the electric force acting on the liquid-vapor interface is very large around the edges of the pillar tops. According to Eq. (12), i.e., $\mathbf{F}_e = -0.5\mathbf{E}\cdot\mathbf{E}\nabla\varepsilon$, the magnitude of the electric force is determined by the module of the electric field intensity $|\mathbf{E}|$ and the gradient of the dielectric permittivity $\nabla\varepsilon$. For dielectric fluids, the permittivity of the liquid phase is larger than that of the vapor phase. Thus, it can be expected that the electric force mainly acts on the liquid-vapor interface since $|\nabla\varepsilon|$ is highest at the liquid-vapor interface. Figure 8 depicts the local distribution of the module of the electric field intensity around the pillar in the case of $\Delta T \approx 0.0169$ and $E_0 = 0.15$ at $t = 11000\delta_t$. As seen in Fig. 8, the maximum value of $|\mathbf{E}|$ appears around the edges of the pillar, which can explain why in Fig. 7 the electric force acting on the liquid-vapor interface is very large around the edges of the pillar tops.

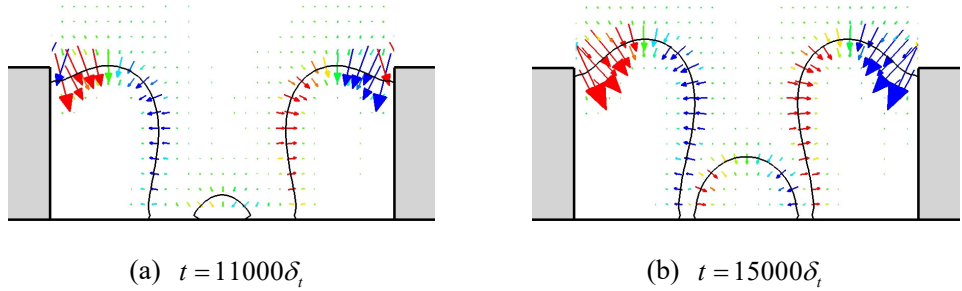

(a) $t = 11000\delta_t$  (b) $t = 15000\delta_t$

**Fig. 7.** Local distributions of the electric force in the case of $\Delta T \approx 0.0169$ and $E_0 = 0.15$ at $t = 11000\delta_t$ and $t = 15000\delta_t$, respectively.

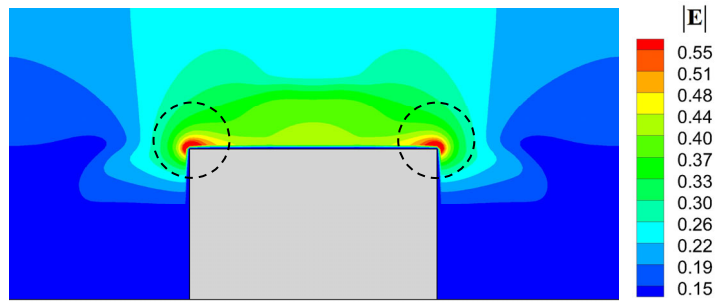

**Fig. 8.** Local distribution of the module of the electric field intensity, namely $|\mathbf{E}|$, around the pillar in the case of $\Delta T \approx 0.0169$ and $E_0 = 0.15$ at $t = 11000\delta_t$.

Owing to the large electric force around the edges of the pillar tops, the movement of the bubbles in the channels will be inhibited by the electric force when they approach the edges of the pillar tops. As a result, the bubbles generated in the channels are unable to cross the edges of the pillar top for growth and departure. Hence, they will coalesce with their adjacent bubbles in each channel and the channels may be blocked by the coalesced bubbles, as seen in Fig. 6(b), which leads to the deterioration of pool boiling heat transfer in the regime II. Such a phenomenon is consistent with the field trap effect discussed in some experiential studies [59, 60]. The present numerical investigation clearly reveals that the negative effect of the electric field on the pool boiling heat transfer mainly arises from the non-uniform distribution of the electric field intensity caused by the surface structure. However, in what follows we will show that the positive effect of the electric field also results from the non-uniform distribution of the electric field intensity.

Figure 9 displays some snapshots of the boiling processes at $\Delta T \approx 0.0252$ (falling into the regime III) in the cases of $E_0 = 0$, $0.15$, and $0.20$, respectively. As seen in the figure, for the case without the electric field (i.e., $E_0 = 0$), the bubbles on the heating surface expand rapidly at such a high wall superheat and the whole surface is quickly covered by a continuous vapor film, although the vapor film may be broken during the bubble necking process. Conversely, when the electric field is applied, the heating surface will not be covered by a continuous vapor film despite the channels may be fully covered by the vapor. Moreover, it is observed that increasing the electric field intensity can promote the departure of the bubbles on the pillar tops. By comparing the third panels of Figs. 9(a), 9(b) and 9(c), we can see that the bubble necking in the cases of $E_0 = 0.15$ and $E_0 = 0.20$ appears much earlier than that in the case of $E_0 = 0$ Accordingly, the bubble departure process can be considerably accelerated for the bubbles nucleated on the pillar tops and the diameter of the departed bubbles can be reduced, which can be seen by comparing the right-hand panels of 9(b) and 9(c) with that of Fig. 9(a).

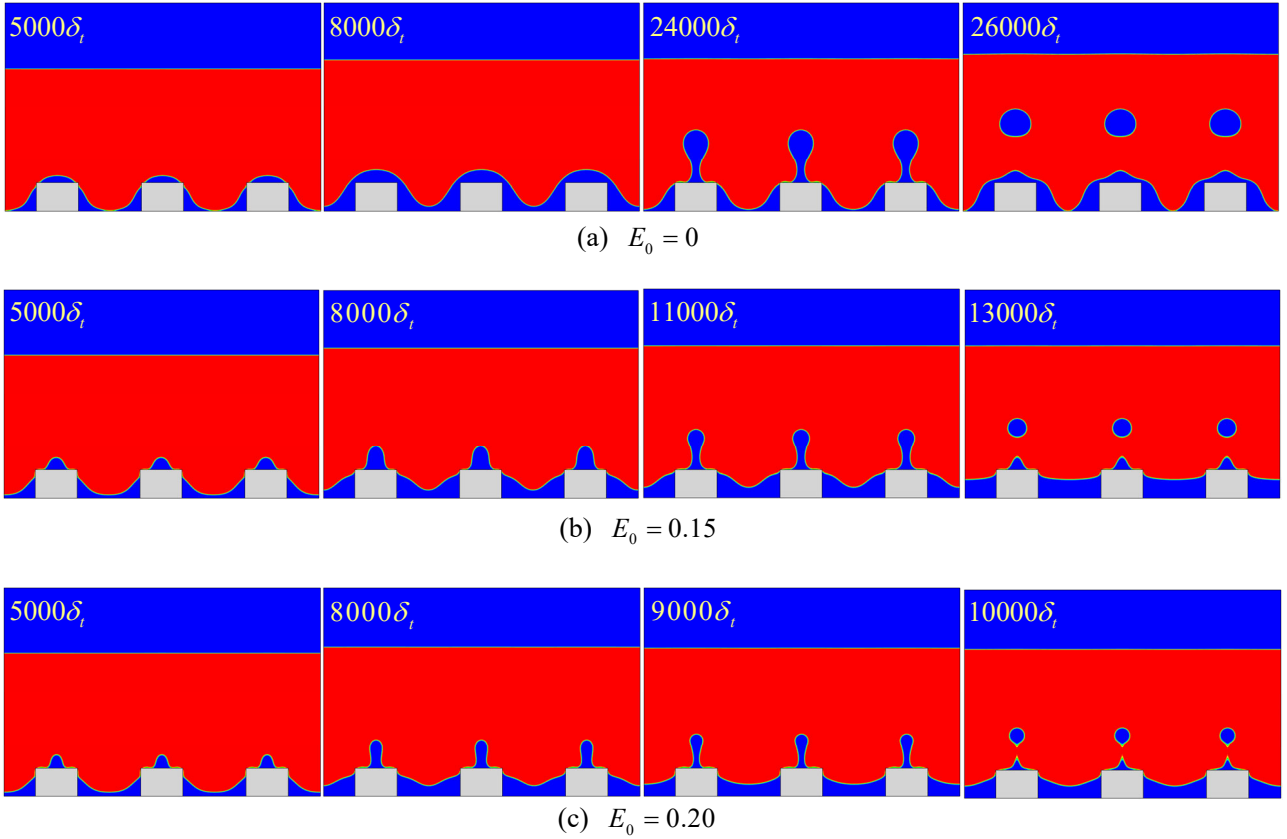

**Fig. 9.** Snapshots of the boiling processes on the pillar-structured surface at $\Delta T \approx 0.0252$ in the cases of (a) $E_0 = 0$, (b) $E_0 = 0.15$, and (c) $E_0 = 0.20$, respectively.

Figure 10 illustrates the local distribution of the electric force in the case of $\Delta T \approx 0.0252$ and $E_0 = 0.15$, which corresponds to the boiling process shown in Fig. 9(b). From Fig. 10 we can see that the large electric force around the edges of the pillar tops can prevent the bubbles in the channels from crossing the edges of the pillar tops, which therefore suppresses the formation of a continuous vapor film covering the whole heating surface and therefore can delay the occurrence of boiling crisis. Meanwhile, the electric force also suppresses the spreading of the bubbles on the pillar tops since the electric force points towards the interior of the bubbles. In particular, during the bubble necking process, the electric forces acting on the bubble neck in the opposite directions can play a *pinching-off* role in the breakup of the bubble neck, as clearly seen in Fig. 10(c). As a result, the departure of the bubbles on the pillar tops can be significantly accelerated. Obviously, such positive effects of the electric field on the pool boiling are also related to the non-uniform distribution of the electric field intensity around the pillar (e.g., see Fig. 8).

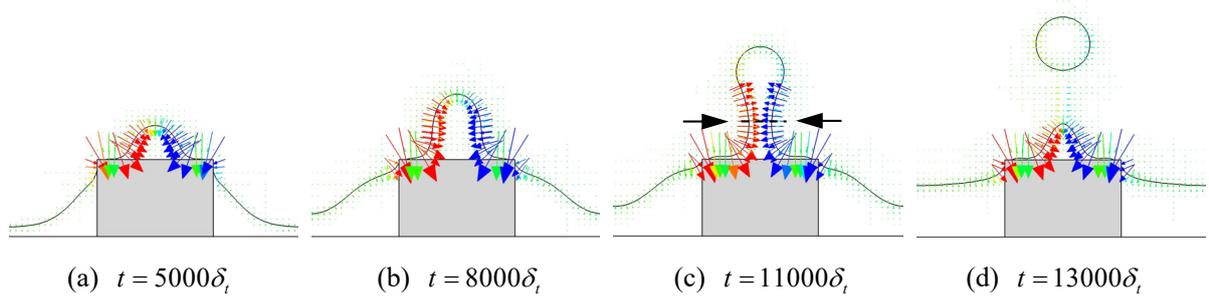

(a) $t = 5000\delta_t$     (b) $t = 8000\delta_t$     (c) $t = 11000\delta_t$     (d) $t = 13000\delta_t$

**Fig. 10.** Local distribution of the electric force in the case of $\Delta T \approx 0.0252$ and $E_0 = 0.15$.

According to the above analysis and discussion, it can be concluded that applying an external electric field causes both positive and negative effects on the pool boiling of dielectric fluids on pillar-structured surfaces. On the one hand, the electric field results in the bubble coalescence in the channels and deteriorates the boiling heat transfer in the regime II. On the other hand, it prevents the coalescence between the bubbles in the channels and those on the pillar tops, which thus suppresses the formation of a continuous vapor film and delays the occurrence of boiling crisis. Moreover, the electric forces acting on the bubble neck can play a *pinching-off* role in the breakup of the bubble neck, which significantly promotes the bubble departure on the pillar tops.

**3.3. Synergistically enhanced pool boiling on pillar-structured surfaces by combining the electric field effect and the effect of mixed wettability**

Based on the boiling mechanism revealed above, in this section we apply mixed wettability to the pillar tops of the pillar-structured surface in order to take the advantage of the positive effect of the electric field in promoting the bubble departure on the pillar tops. Figure 11 illustrates the pillar-structured surface with mixed wettability, where the wettability-modified regions are uniformly distributed on each pillar top. For the pillar-structured surface with the pillar width of $W = 60$ l.u. and $W = 100$ l.u., the number of the wettability-modified regions is chosen as 3 and 5, respectively. The width of the wettability-modified regions is taken as $W_{\text{mod}} = 6$ l.u. and the center-to-center spacing between two wettability-modified regions is chosen as $D_s = 20$ l.u. Moreover, the wettability-modified regions are hydrophobic with the contact angle of $\theta_{\text{pho}} \approx 93.6°$, while the other regions remain hydrophilic with the contact angle of $\theta_{\text{phi}} \approx 36.6°$.

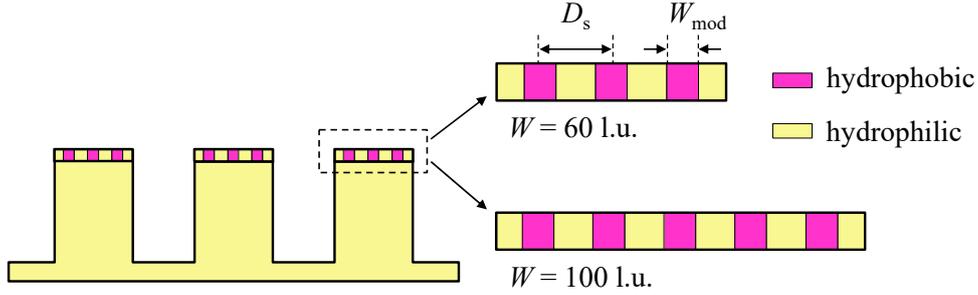

**Fig. 11.** Schematic of the pillar-structured surface with mixed wettability, where the wettability-modified regions are uniformly distributed on each pillar top.

Figure 12 compares the boiling performance between the pillar-structured surface with mixed wettability and the pillar-structured surface with homogeneous wettability under an electric field when the pillar width is taken as $W = 60$ l.u., including the boiling curves and HTCs. For convenience, the pillar-structured surface with mixed wettability and the pillar-structured surface with homogeneous wettability are respectively called the mixed surface and the base surface in the rest of the present paper. Firstly, a comparison is made in the case of $E_0 = 0.15$ between the mixed surface and the base surface. From Fig. 12(a) we can see that, under the electric field with $E_0 = 0.15$, the boiling curve of the mixed surface is shifted to the left of that of the base surface. Accordingly, an upward shift of the HTC curve can be observed in Fig. 12(b) for the mixed surface in comparison with that of the base surface under the electric field with $E_0 = 0.15$. Similar results can also be observed when the electric field intensity is increased to $E_0 = 0.25$, which implies that the boiling heat transfer on the pillar-structured surface under the electric field has been further enhanced by incorporating the effect of mixed wettability. For instance, compared with those of the base surface under the electric field with $E_0 = 0.15$ and 0.25, the maximum HTCs of the mixed surface have been increased by 21% and 15.7%, respectively.

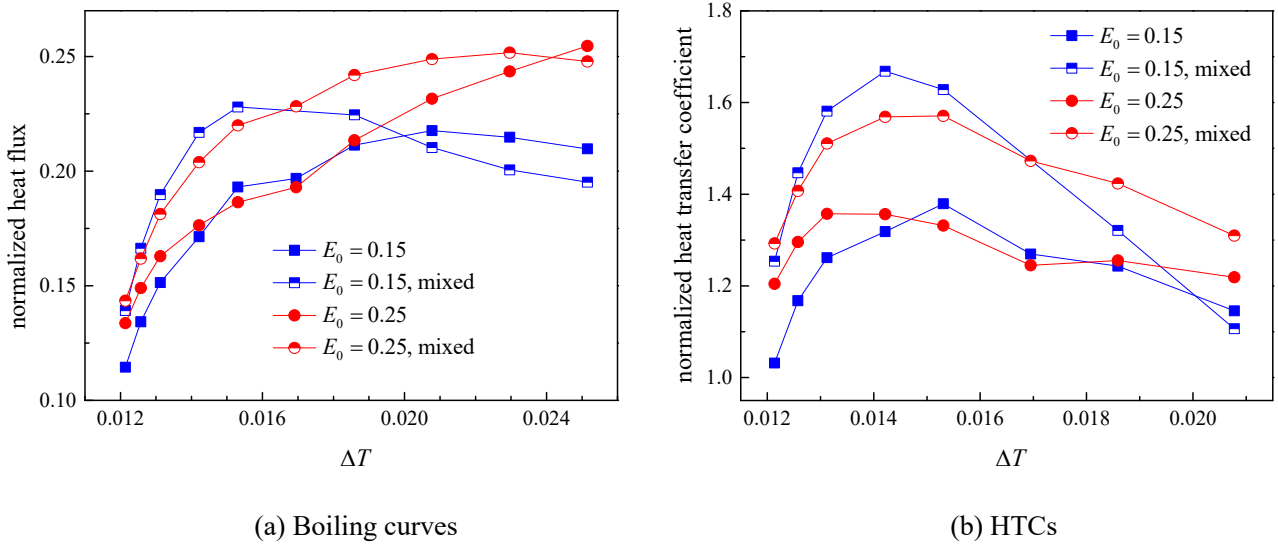

(a) Boiling curves  (b) HTCs

**Fig. 12.** Comparisons of the boiling curves and HTCs between the pillar-structured surface with mixed wettability (mixed surface) and the pillar-structured surface with homogeneous wettability under an electric field ($W = 60$ l.u.).

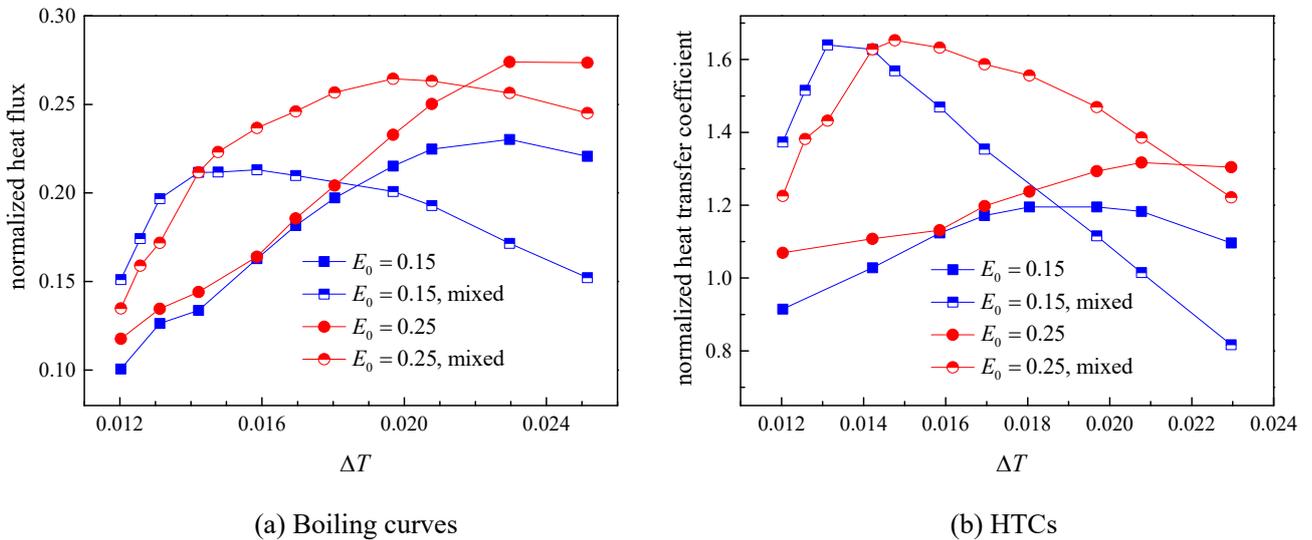

(a) Boiling curves  (b) HTCs

**Fig. 13.** Comparisons of the boiling curves and HTCs between the pillar-structured surface with mixed wettability (mixed surface) and the pillar-structured surface with homogeneous wettability under an electric field ($W = 100$ l.u.).

Furthermore, when the pillar width is increased to $W = 100$ l.u., similar phenomena can also be observed in Fig. 13, which also shows that the boiling curve of the mixed surface is shifted to the left of that of the base surface and the maximum HTC can be considerably improved. Particularly, it is observed that the boiling performance on the pillar-structured surface can be enhanced synergistically with the CHF being increased by imposing the electric field and the maximum HTC being improved by applying the mixed wettability. For example, according to Figs. 13 and 3, the CHF and $HTC_{max}$ of the mixed surface under the electric field with $E_0 = 0.25$ are increased by 27.1% and 27.8%, respectively, in comparison with those of the base pillar-structured surface in the absence of the electric field.

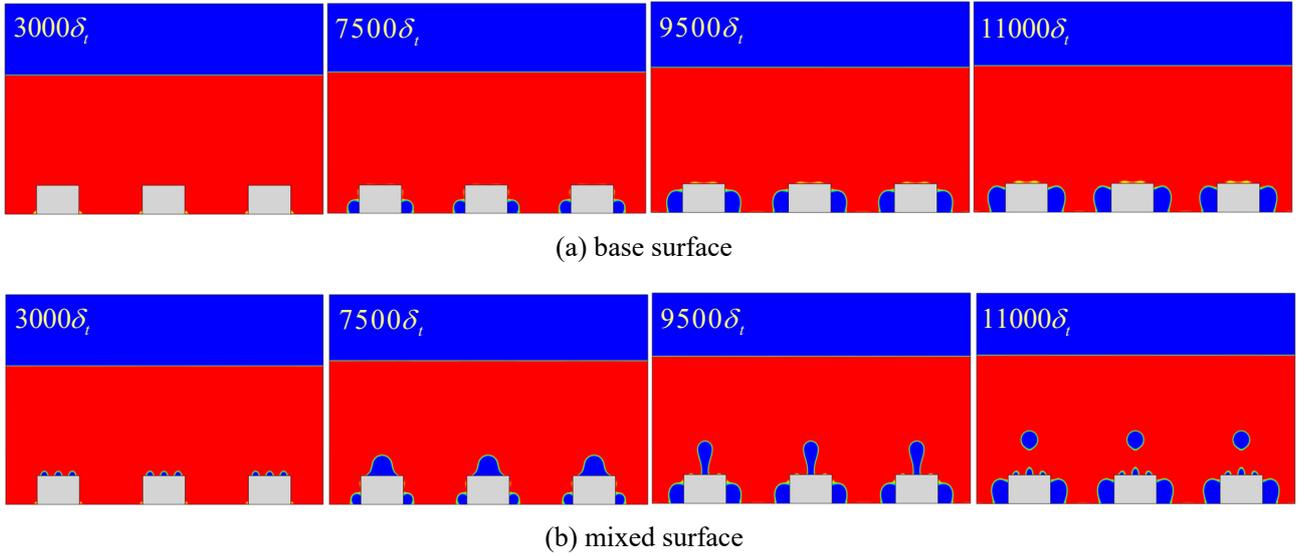

(a) base surface

(b) mixed surface

**Fig. 14.** Snapshots of the boiling processes on (a) the base surface and (b) the mixed surface at $\Delta T \approx 0.0153$ under the electric field with $E_0 = 0.15$.

By taking the case of $W = 60$ l.u. as an example, Fig. 14 compares the bubble dynamics during the boiling process at $\Delta T \approx 0.0153$ between the base surface and the mixed surface under the electric field with $E_0 = 0.15$. From the figure we can see that, for the base surface, at $\Delta T \approx 0.0153$ the bubbles are only nucleated in the channels and no bubbles are observed on the pillar tops of the base surface. However, for the mixed surface, the bubbles are not only generated in the channels but also nucleated at the wettability-modified regions on the pillar tops, as seen in Fig. 14(b). The bubble nucleation on the pillar tops of the mixed surface is promoted by the wettability-modified regions with hydrophobicity, which reduces the energy barrier of the liquid-vapor phase change and therefore increases the density of the nucleation site. As time goes by, the bubbles nucleated on each pillar top will merge into a larger bubble, which can be seen in the second panel of Fig. 14(b). Subsequently, the bubble will shrink inward rapidly owing to the electric force imposing on the liquid-vapor interface, as seen in the second panel of Fig. 15, in which the local distribution of the electric force is displayed. After that, the electric force will play a pinching-off role during the bubble necking process for promoting the bubble departure, which can be seen in Figs. 15(c) and 15(d). To sum up, the effect of mixed wettability and the positive effect of the electric field have been combined to generate more bubbles and then to promote their departure for enhancing the boiling heat transfer.

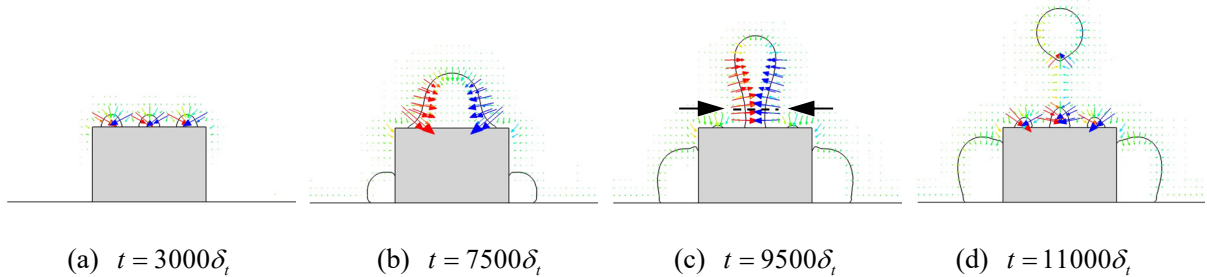

(a) $t = 3000\delta_t$  (b) $t = 7500\delta_t$  (c) $t = 9500\delta_t$  (d) $t = 11000\delta_t$

**Fig. 15.** Local distribution of the electric force in the case of $\Delta T \approx 0.0153$ and $E_0 = 0.15$ for the mixed surface.

Figure 16 compares the normalized transient heat fluxes on the pillar tops between the base surface and the mixed surface at $\Delta T \approx 0.0153$ under the electric field with $E_0 = 0.15$. From the figure it can be seen that the normalized transient heat flux on the pillar tops of the mixed surface is always higher than that of the base surface during the boiling process. Quantitatively, the average value of the normalized transient heat flux on the pillar tops of the mixed surface is about 1.79 times of that of the base surface, which means that the heat flux on the pillar tops of the pillar-structured surface under the electric field with $E_0 = 0.15$ has been increased by about 79% by applying the mixed wettability to the pillar tops.

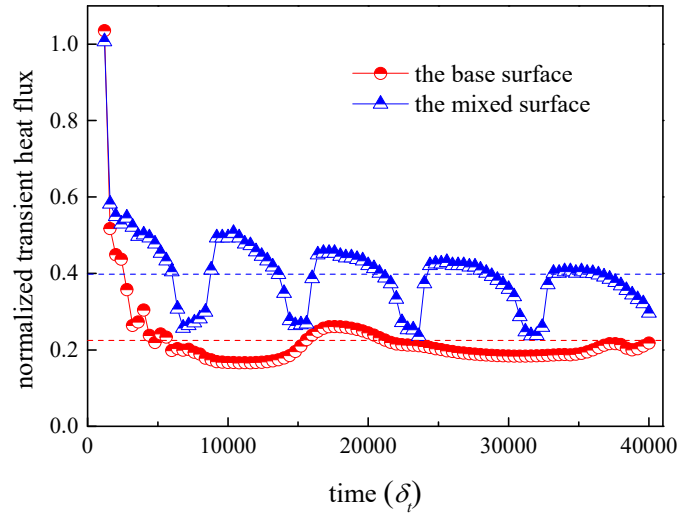

**Fig. 16.** Variations of the normalized transient heat fluxes on the pillar tops of the base surface and the mixed surface at $\Delta T \approx 0.0153$ under the electric field with $E_0 = 0.15$.

## 4. Conclusions

In recent years, considerable attention has been paid to the enhancement of pool boiling of dielectric fluids by applying an electric field. However, the boiling characteristics and the associated enhancement mechanism of pool boiling under an electric field have not been well understood. In this paper, by using a phase-change LB model coupled with an electric field model, we have numerically investigated the boiling performance of dielectric fluids on pillar-structured surfaces under an electric field, with the associated boiling mechanism being analyzed. The present numerical investigation reveals that applying an electric field causes both positive and negative effects on the pool boiling on pillar-structured surfaces. It is found that, under the action of an electric field, the electric force prevents the bubbles nucleated in the channels from crossing the edges of the pillar tops. On the one hand, such an effect results in the bubble coalescence in the channels and blocks the paths of liquid supply for the channels, which leads to the deterioration of pool boiling in the medium-superheat regime. On the other hand, it prevents the coalescence between the bubbles in the channels and those on the pillar tops, which can suppress the formation of a continuous vapor film covering the whole heating surface and therefore delay the occurrence of boiling crisis. Moreover, it has been clearly shown that the electric force acting on the liquid-vapor interface can play a *pinching-off* role during the bubble necking process, which can significantly promote the bubble departure on the pillar tops. Based on the revealed boiling mechanism, wettability-modified regions have been applied to the pillar tops of the pillar-structured surface to further enhance the boiling heat transfer under the electric field. The numerical results show that the boiling performance on the pillar-structured surface can be enhanced synergistically with the CHF being increased by imposing an electric field and the maximum HTC being improved by applying mixed wettability to the pillar-structured surface.

## Acknowledgments

This work was supported by the National Natural Science Foundation of China (Grant No. 52176093 and No. 51822606).